\documentclass[10pt,letterpaper]{article}
\usepackage[top=0.85in,left=1.5in,footskip=0.75in,marginparwidth=2in]{geometry}

\usepackage{microtype}
\usepackage{graphicx}
\usepackage{subfigure}
\usepackage{booktabs} % for professional tables
\usepackage{soul}
\usepackage{amssymb}
\usepackage{amsmath}
\DeclareMathOperator*{\argmin}{argmin}
\usepackage{hyperref}
\usepackage{natbib}

\usepackage[utf8]{inputenc}

\usepackage{cite}

\usepackage{nameref,hyperref}

\usepackage[right]{lineno}

\usepackage{microtype}
\DisableLigatures[f]{encoding = *, family = * }

\usepackage{changepage}

\usepackage[aboveskip=1pt,labelfont=bf,labelsep=period,singlelinecheck=off]{caption}

\makeatletter
\renewcommand{\@biblabel}[1]{\quad#1.}
\makeatother

\usepackage{lastpage,fancyhdr,graphicx}
\usepackage{epstopdf}
\fancyheadoffset[L]{2.25in}
\fancyfootoffset[L]{2.25in}

\usepackage{color}

\definecolor{Gray}{gray}{.25}

\usepackage{graphicx}

\usepackage{sidecap}

\usepackage{wrapfig}
\usepackage[pscoord]{eso-pic}
\usepackage[fulladjust]{marginnote}
\reversemarginpar

\begin{document}
\vspace*{0.35in}

% title goes here:
\begin{flushleft}
{\Large
\textbf\newline{Don't stop the training: continuously-updating self-supervised algorithms best account for auditory responses in the cortex}
}
\newline
% authors go here:
\\

Pierre Orhan\textsuperscript{1,*},
Yves Boubenec\textsuperscript{1},
Jean-Remi King\textsuperscript{1,2},
\\
\bigskip
\bf{1} École normale supérieure, PSL University, CNRS, Paris, France
\\
\bf{2} Facebook AI Research, Paris, France
\\
\bigskip
* orhan.pierre.france@gmail.com

\end{flushleft}

\begin{abstract}

% Previous works showed that activities of neural networks responding to specific inputs are predictive of brain responses to the same inputs. We report similar results for recording of ferret auditory cortex predicted from the activity of a Wav2vec2 network. We use the data from {\color{blue} cite Agnes }, Ferret repeatedly heard long sequences of 10 seconds sounds, while their auditory cortex was recorded using functional ultrasound. Surprisingly, we observe that by letting the network update itself continuously through the same sensory experiment this predictability is increased. This increase was absent when the network could update itself through the same but differently ordered sequence of sound. Additionally, it did not reflect an increase of similarity between the states of real and artificial networks. Indeed, if measured after this sequence of self-supervised learning, the network activity to the set of sound is as predictive of brain activity as before. Instead, it was necessary to extract activities during the hearing sequence, from ephemeral networks, who learned based on the ordered previous context. Therefore, backpropagation through an artificial neural network was not associated with long-term plastic change in the ferret brains but rather a contextual adaptive effect. Together, these results build a link between backpropagation in a neural network and adaptation in the ferret auditory cortex. They show the importance of providing normative models with self-supervised capabilities and the sensory history of experimental subjects.

Over the last decade, numerous studies have shown that deep neural networks exhibit sensory representations similar to those of the mammalian brain, in that their activations linearly map onto cortical responses to the same sensory inputs. However, it remains unknown whether these artificial networks also \emph{learn} like the brain. To address this issue, we analyze the brain responses of two ferret auditory cortices recorded with functional UltraSound imaging (fUS), while the animals were presented with 320 10\,s sounds. We compare these brain responses to the activations of Wav2vec 2.0, a self-supervised neural network pretrained with 960\,h of speech, and input with the same 320 sounds. Critically, we evaluate Wav2vec 2.0 under two distinct modes: (i) "Pretrained", where the same model is used for all sounds, and (ii) "Continuous Update", where the weights of the pretrained model are modified with back-propagation after every sound, presented in the same order as the ferrets. Our results show that the Continuous-Update mode leads Wav2Vec 2.0 to generate activations that are more similar to the brain than a Pretrained Wav2Vec 2.0 or than other control models using different training modes. These results suggest that the trial-by-trial modifications of self-supervised algorithms induced by back-propagation aligns with the corresponding fluctuations of cortical responses to sounds. Our finding thus provides empirical evidence of a common learning mechanism between self-supervised models and the mammalian cortex during sound processing.

\end{abstract}

\section{Introduction}
\label{submission}

\paragraph{General challenge}
Over the last decade, numerous studies have shown striking similarities between deep neural networks and the mammalian brain. In particular, deep convolutional neural networks pretrained on visual \citep{guclu_deep_2015,yamins_performance-optimized_2014,eickenberg_seeing_2017} and auditory categorization  \citep{kell_task-optimized_2018,millet_inductive_2021} have been shown to elicit representations that linearly map onto those of the cortex in response to the same input. These results suggest that the computational solution found by these algorithms is partially similar to the computations implemented in sensory cortices.

%\item Challenge: 
However, this functional convergence highlights a major challenge: unlike these algorithms, the brain is rarely fed with supervision signals. Rather, it must largely rely on a training objective that is independent of external feedback -- an approach referred to as unsupervised or self-supervised learning \citep{becker_self-organizing_1992,bromley_signature_1993,mikolov_efficient_2013,devlin_bert_2019,joulin_bag_2016}.
%
% Self-supervised learning rely on the exploitation of statistical properties of the data received by the networks. An example of such statistical property are the contextual dependencies of different information.
Remarkably, neural populations modulate their responses depending on variably distant contexts %maintain contextual information for long time periods, and can modulate the responses to subsequent stimuli
\citep{nikolic_distributed_2009,klampfl_quantitative_2012}. These contextual modulations might therefore %play a role in biological 
reflect self-supervised  mechanisms in the brain.

% Second, the brain is continuously learning to efficiently adapt to its changing environment (XXX review bayesian litterature): \emph{i.e.} the brain is almost not likely constrained to separate training and inference modes of working (XXXX): demonstration of 1-shot learning are commonplaces in XXX, XXX and XXX.

\paragraph{Hypothesis}
Here, we hypothesize that self-supervised models that continuously update their internal parameters after each input make better models of the brain than their pretrained and "frozen" counterparts. To test this hypothesis, we focus on a state-of-the-art self-supervised algorithm: Wav2Vec 2.0. We then apply a standard similarity analysis \citep{yamins_using_2016} to compare its activations to those of brain two ferrets recorded with Functional UltraSound (fUS) by \citep{landemard_distinct_2021}, while the animals listened to multiple repetitions of 320 distinct 10\,s-long sounds.

\paragraph{Contributions} Our results show for the first time that continuously-updating a deep neural network makes it more brain-like than using its "frozen" and pretrained counterpart. This finding delineates a clear path to model self-supervision mechanisms in the brain.

\begin{figure*}[ht]
% \vskip 0.2in
\vskip 0.1in
\centering
% \begin{center}
\centerline{\includegraphics[width=\linewidth]{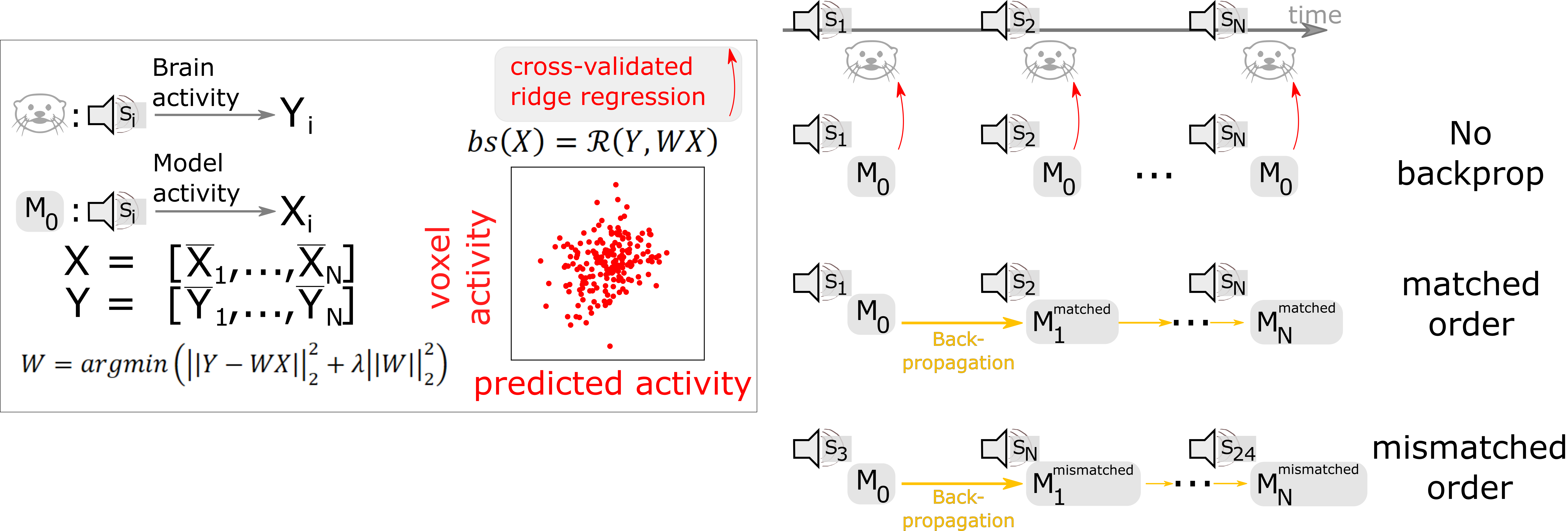}}
\caption{Schematic view of the experiment. Representation similarities between model and brain responses to sounds are estimated by a cross-validated ridge regression. Model listened to each sounds either without modification or by continuously training themselves. Sound were either presented in the same (matched) order as the Ferret listened or a wrong (mismatched) order.}
\label{fig:figIntro}
% \end{center}
% \vskip -0.2in
\vskip -0.1in
\end{figure*}

% Neural populations in sensory cortices carry information about preceding stimuli for several hundred of ms, a process referred to as iconic memory in the visual system. This memory persists throughout epochs without stimulus, and even when new stimuli are presented (Nikolic biology 2009). In the auditory system, very similar population dynamics have been reported (Klampfl 2012) and are postulated to be at the core of echoic memory. In turn, deep neural networks, and in particular self-supervized, do retain a trace of preceding stimuli (REF). However, to date, it is unknown whether this short-term learning will impact artificial and biological networks in the same way. We thus propose to test whether preceding context affected responses to upcoming stimuli in a comparable manner between a self-supervized neural networks and ferret auditory cortex both presented with identical sequences of natural sounds.

\section{Methods}

We aim to compare the activations of a deep neural networks to those of the ferret auditory cortices in response to the same sounds.

\subsection{Wav2vec 2.0 responses to sounds}

Wav2Vec 2.0 is a deep neural network composed of two main parts: (i) an encoder, which consists of a hierarchy of convolutional layers and (ii) a decoder, which consists of a non-causal transformer. These two encoder/decoder modules are trained end-to-end to predict masked segments of the encoded signals given their surrounding context. % In parallel, Wav2Vec 2.0 recognizes the occurrence of these abstract vocabulary in the audio signal. Together these two objectives mathematically translate to a self-supervised objective, which error is backpropagated through the network to update its parameters.
%The Wav2Vec 2.0 architecture is originally pre-trained on a speech dataset. We use this pre-trained model and extend its training using the 320 sounds.

% The model activations are compared to fUS responses. Functional UltraSound records blood flow as a proxy for brain activities. Blood velocity is estimated by the Doppler frequency shift of sound waves reflecting off blood particles. fUS is similar to fMRI in that it uses particle flow as a proxy for neuronal activity but it has an improved temporal () and spatial (100 $\mu$m) resolution. To have an order of magnitude, voxel sizes correspond to the spatial integration range of fast-spiking interneurons.

% hyper parameters
We use the Huggingface \footnote{https://huggingface.co/} implementation of the Wav2vec 2.0   \citep{baevski_wav2vec_2020}. This model was originally trained in \citep{baevski_wav2vec_2020} on the Librispeech dataset (960\,h) with 7 convolutional layers (each with 512 units), 12 transformer blocks and 102.4k quantizations.
To investigate the similarities between brain responses and Wav2Vec 2.0, we focus on the representation of the encoder, which, like the brain (but unlike the decoder) are causally defined. 

We analyze Wav2Vec 2.0 under three distinct modes.

First, we use the pretrained Wav2Vec 2.0: $M_0$. To this aim, we input $M_0$ with each of the sounds provided by \citep{landemard_distinct_2021} and sampled at 16\,kHz. We then extract the mean temporal activity $\bar{X}_{i}\in \mathbb{R}^{d}$ of the $d=7\times512=3,584$ artificial neurons of the encoder in response to each sound $s_{i}$.

\begin{equation}
    \bar{X}_{i} = \frac{1}{T} \sum_{t=1}^T M_0(s_i,t)
\end{equation}

We refer to the activations of this pretrained ("frozen") model as "no backprop" activations.

Second, we follow the same procedure on a Wav2Vec 2.0 continuously updated after  each sound. Specifically, let us label the $N$ sounds $s_1,...,s_N$, played in a fix order to the ferret ($N =200$ or $N=120$). Starting from an initial pretrained model $M_0$, we extract the responses of the encoder to sound $s_1$ and average them over the 10\,s time window: $\bar{X}_1$.

We then estimate the  self-supervised contrastive loss to this sound using a random binary mask $u\in \{0,1\}^T$ (following the default parameters provided by \citep{baevski_wav2vec_2020}). This prediction error $\mathcal{L}\left(s_{i}, M_{i-1}, m\right)$ is back-propagated to the network and weights are updated by an Adam optimizer, with default parameters and learning rate $3\times10^{-6}$. We then use this updated model $M_1$ to extract the activations of following sound $s_2$: $M_1(s_2)=\bar{X}_2$ and iterate the process until all sounds have been presented to the network. 

%Generally, we write:

\begin{equation}
    \bar{X_i} = \frac{1}{T} \sum_{t=1}^T M_{i-1}(s_i,t)
\end{equation}
\begin{equation}
    M_{i} \xleftarrow{  } \text{Adam}(M_{i-1},\mathcal{L}\left(s_{i}, M_{i-1}, m\right)) %\mathcal{L}(M_i(s_{i-1}), M_i(u\times s_{i-1}))
\end{equation}
\color{black}
Random masks add stochasticity to the back-propagation. To minimize this issue, we repeated each this process across three random seeds and average across seeds.

We refer to the resulting activations as "Backprop in matched order" activations.

Third, we replicate the same procedure as the Continuous Update mode, but use a random sound order. We refer to the resulting activations as "Backprop in mismatched order".

\subsection{Cortical responses to sounds}

Functional UltraSound (fUS) imaging is a technique that captures the hemodynamic responses of the brain with a very high spatial resolution and spatial coverage (100\,µm in plane, several cm in depth). Similar to functional Magnetic Resonance Imaging (fMRI), fUS imaging captures slow signals and local variations in blood volumes, that correlate with the activity of local neuronal populations \citep{mace_functional_2011,bimbard_multi-scale_2018,landemard_distinct_2021}. %We have been adapting this neuroimaging technique to study the functional organization of auditory cortex in the awake ferret (BIMBARD 2018). 
%More recently, we developed a cross-species approach to reveal the unique contribution of the human auditory cortex in processing speech and music higher-order acoustics statistics (LANDEMARD 2021). 

%\item 
We use the preprocessed fUS data recorded by \citep{landemard_distinct_2021}, in which two ferrets, 'F1' and 'F2', passively listened to 320 distinct sounds. % over two experiments ('N' with 200 and 'V' with 120 sounds). 
%Experiments are repeated for recording of the two hemisphere ('R' right or 'L' left). %In each hemisphere a day of recording is dedicated to acquire each slice of cortex. 
%
These 320 sounds can be categorized in four major categories: (i) ferret vocalization, (ii) speech, (iii) music, and (iv) other environmental sounds.
Each trial lasted 20\,s and consisted of 10\,s of silence and  10\,s of sounds. Trials were separated by a silence period randomly varying from 3 to 5\,s.
Each day of recording, the two ferrets listened to 4 blocks composed of the same 10\,s sounds ($N=200$ or $N=120$ depending on the day). Critically, the order of the sounds is specific to each block. This experimental design implies that we can model brain responses to individual sounds depending on their long-distant context (i.e. $>$10\,s).
Each brain "slice"consists of an average of $920\pm163$ voxels and was recorded on a distinct day.

F1 and F2 were recorded over 52 and 73 days, respectively, for a total of 32 and 45 recorded slices ($\approx$ 74,459 voxels), respectively. 

The preprocessing of the data is detailed in \citep{landemard_distinct_2021}. In brief, canonical correlation analysis (CCA) was used to remove the spatial components shared across cortical voxels and voxels located outside of the cortex. The resulting data was then normalized by the baseline activity recorded before and after each sound presentation. 

%Secondly, the dimensionality was reduced by finding response components shared between ferret, slices and odd or even repetition of the experiments. 

%In the original paper the data was pre-processed through two stage. First, components shared between out/in of cortex voxels were removed by CCA and the activity was normalized by the baseline activity (the fus response was recorded for a few seconds before and after sound presentation). Secondly, the dimensionality was reduced by finding response components shared between ferret, slices and odd or even repetition of the experiments. For the analysis carried here we focused on part of the response specific to each trial. Therefore we used the dataset after pre-processing with the first stage of the pipeline only. This explains the relatively small values of brain scores reached. If one computes brain scores with the fully denoised dataset (projected on a common space of low dimensionality), we observed the brain scores to be larger, for example distributed around 0.6 when the dimensionality is 8.
%% way to improve: note that other paper have the same effect size

\subsection{Brain Scores: quantifying the similarity between model and brain activations}

% Multiple studies have used this metric to quantify the degree of similarity between signal processing in brains and deep neural networks. {\color{blue} cite classical papers} This study aims at discovering subtle aspect of this similarity, and observe if continuous update change it.

A brain score summarizes the representational similarity between a deep learning algorithm and brain responses \citep{yamins_performance-optimized_2014}. Here, the brain score measures the maximal correlation between a linear combinations of artificial neurons and the activity of a unique voxel. 

Because the hemodynamic responses are notoriously slow, we compute brain scores on activations averaged over the 10\,s-long time period of each stimulus. Brain scores are here computed using a stratified cross-validated ridge regression.
Specifically, for each train split $I$, we fit a $l2$-regularized 'ridge' regression to predict the fUS activity in a given voxel $v$ in response to sound $i \in I$ : $\bar{y}_i \in \mathbb{R}$ from a linear combination $w \in \mathbb{R}^{d}$ of the model activations $\bar{X}_i$ in response to the same sound and averaged over the 10\,s time samples:

\begin{equation}
    \argmin_w \sum_{i} (\bar{y}_i - w\bar{X}_i)^2 + \lambda ||w||^2
\end{equation}

with $\lambda$, the regularization hyper-parameter. We use the `RidgeCV` estimator provided by scikit-learn\footnote{https://scikit-learn.org/} to efficiently fit this ridge regression with an internal grid search to identify the optimal $\lambda$ (for each voxel) over a logarithmically spanned space of 40 values ranging from $10^{-8}$ to $10^{8}$. We normalize (zero-mean, unit-variance) the activations of the voxels and of the artificial neurons, by estimating the means and standard deviation separately for each training and testing fold.

We then predict the mean activity of each voxel in response to each sound $s_j$ in the test set $J$, and summarize the exactness of this model-to-brain mapping with a Pearson correlation $r$

\begin{equation}
    r_v = corr(w \cdot \bar{X}_J, \bar{y}_J)
\end{equation}

Throughout the paper, we report the average brain score $r$ across all voxels recorded from an animal, mapped from the responses of all artificial neurons in the encoder (convolutional layers) of Wav2Vec 2.0. 

The fUS data was subdivided in 4 cross-validation splits, defined in \citep{landemard_distinct_2021}.

To evaluate the robustness of our results, we perform statistics across slices ($sr=36$ for F1 and $sr=45$ for F2) for each animal separately. 
These $sr$ brain scores consist of the average of the $4\times3\times V$ brains scores of each slice, for 4 experimental block repetitions, 3 %in silico training repetition
random seeds and $V$ voxels per slice. Statistical test associated with figures \ref{fig:fig2}, \ref{fig:fig3} and \ref{fig:fig5} are made with these $sr$ statistics using the scipy.stats \footnote{https://scipy.org/} implementation of the Wilcoxon test. The corresponding $sr$ values are independent observations because slices are recorded on different days.

\section{Results}

\subsection{The activations of the pretrained Wav2Vec 2.0 linearly map onto the brain responses to sounds}

We first test whether the activations of a pretrained deep neural network -- here Wav2Vec 2.0 -- linearly predict sound-evoked brain responses recorded with fUS in ferret auditory cortex.
To this aim, we fit a ridge regression to predict the activity elicited by each sound in each voxel from the activations of Wav2Vec 2.0 input with the same sounds. We then summarize the similarity between Wav2Vec 2.0 and the brain by computing a "brain score" \citep{yamins_performance-optimized_2014}, \emph{i.e.} the correlation between the true brain activity, and the activity predicted from the ridge regression.% across all voxels
%, slices, slice 
%and repetitions %and experiments
%for each of the two ferrets.

%,  similarly to what has been recently found in  

\begin{figure}[ht]
% \vskip 0.2in
\begin{center}
\centerline{\includegraphics[width=\columnwidth]{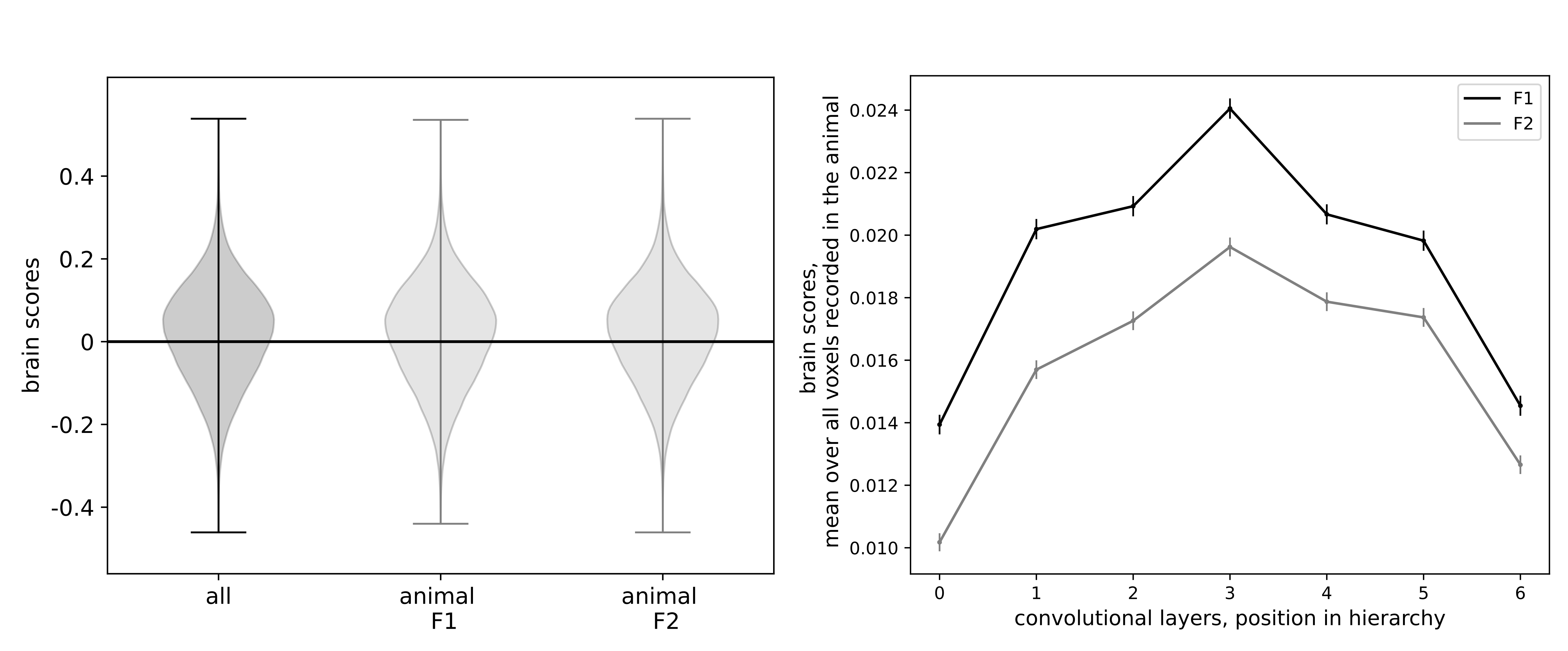}}
\caption{Predicting sound-evoked cortical responses from Wav2vec 2.0 activations. Left, violin plot for the brains scores of all voxels of the animal, concatenated across experiment repetitions. Brain scores are computed using the activity of encoding convolutional layers in a pre-trained Wav2vec 2.0 network. Right, brain scores when using solely the activity of one layer of the encoder.}
\label{fig:fig2}
\end{center}
% \vskip -0.2in
\end{figure}

On average across voxels, the brain scores of a pretrained Wav2Vec 2.0 are significantly distributed above zero for both ferrets (F1: $r = 0.0213\pm 10^{-4}, p< 10^{-7} $, F2: $r = 0.0176 \pm 10^{-4}, p<10^{-8}$, Wilcoxon signed-rank test for mean larger than 0) see Figure \ref{fig:fig2}. This result confirms recent studies demonstrating similarities between pretrained deep neural networks and single cell \citep{yamins_performance-optimized_2014}, magneto-encephalography \citep{caucheteux_language_2020}, and fMRI recordings \citep{caucheteux_gpt-2s_2021, khaligh-razavi_fixed_2017, kell_task-optimized_2018} and extends them to fUS recordings. Remarkably, our effect sizes (up to $r\approx0.5$, and on average across voxels $r\approx0.02$) is similar to what has been found in fMRI \citep{lebel_voxelwise_2021,millet_inductive_2021}, which suggests similar signal-to-noise ratios.

We then compare the ability of each convolutional layer of the Wav2Vec 2.0 encoder to linearly predict the fUS recordings. We observe that intermediate layers are more predictive than deeper layers, in accordance to the intermediate role of the auditory cortex in the sound processing hierarchy \citep{kell_task-optimized_2018, millet_inductive_2021}, and similar to results for visual processing \citep{yamins_performance-optimized_2014,eickenberg_seeing_2017}

\subsection{Continuously updating Wav2Vec 2.0 makes it more brain-like}

\begin{figure*}[ht]
% \vskip 0.2in
% \begin{center}
\centering
\centerline{\includegraphics[width=\linewidth]{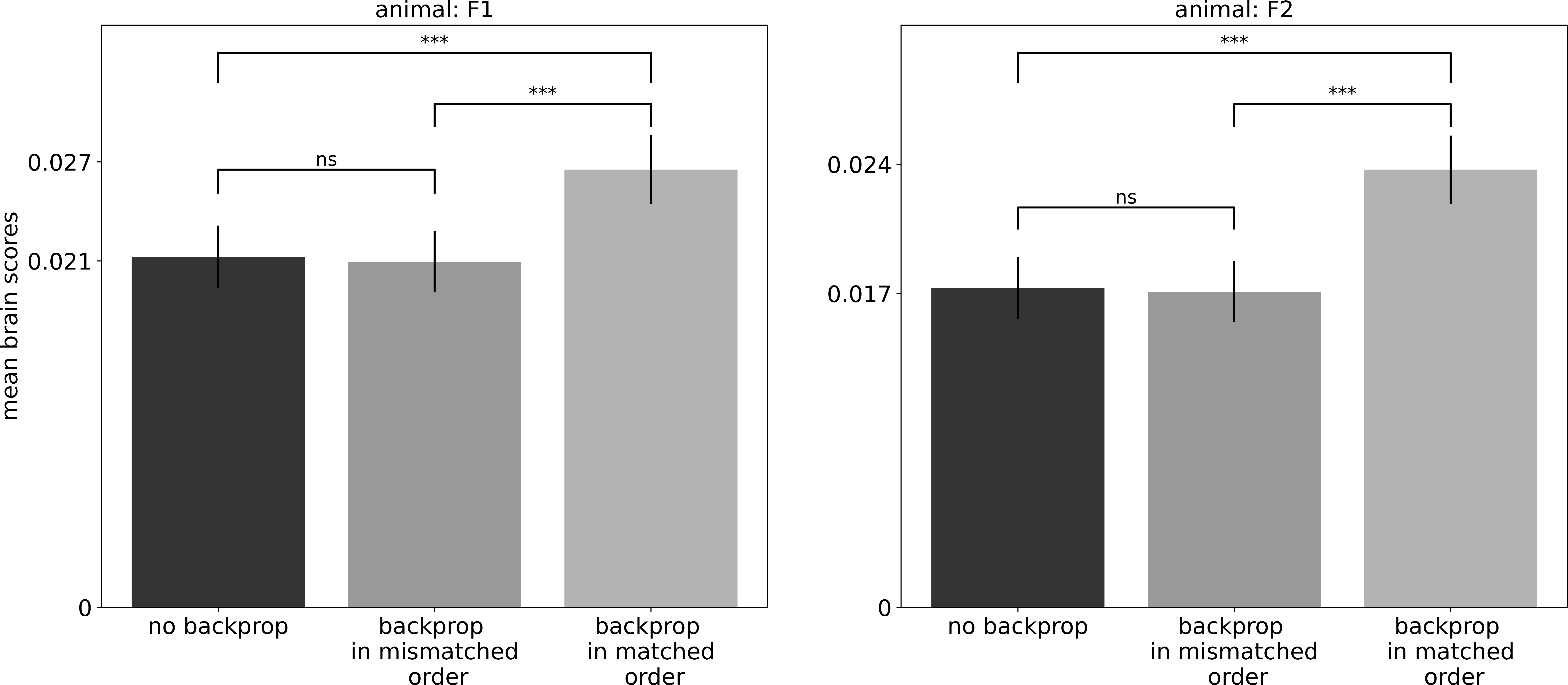}}
\caption{Continuously updated networks with backpropagation are more predictive of brain activity than frozen network. 
We report mean brain scores and standard error of the mean, with either an untrained network (no backprop), networks trained continuously with sounds presented in right order (same order as heard by the Ferret) or wrong order.
}
\label{fig:fig3}
% \end{center}
% \vskip -0.2in
\end{figure*}

% figure in supplementary
% \begin{figure}[ht]
% % \vskip 0.2in
% \begin{center}
% \centerline{\includegraphics[width=\columnwidth]{figures/figure_2bis.png}}
% \caption{Shuffle control: we repeat the experiment with 100 random order and compare the results to 3 run using the correct order but different (random) masking for the unsupervised loss computed at each sound. }
% \label{fig:fig4}
% \end{center}
% % \vskip -0.2in
% \end{figure}

\begin{figure}[ht]
% \vskip 0.2in
\begin{center}
\centerline{\includegraphics[width=\linewidth]{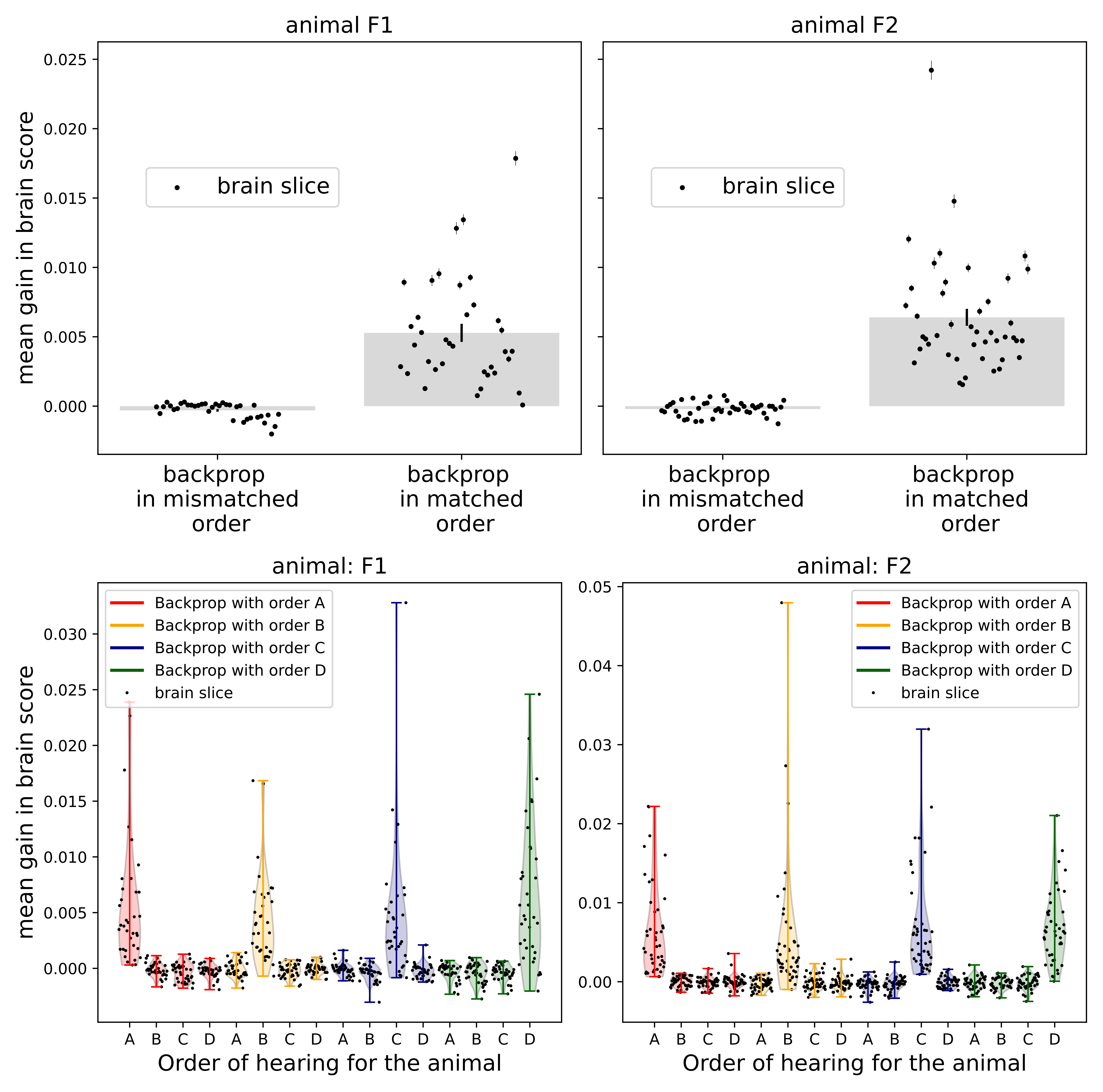}}
\caption{Top, mean and standard error of the mean over all voxels in the different slices are shown. Bottom: each brain recording is repeated 4 times sequentially with different sounds order. Learning with the order of repetition X to predict the activity of a repetition Y (X $\neq$ Y) did not improve the brain score. }
\label{fig:fig4}
\end{center}
% \vskip -0.2in
\end{figure}

%old legend:
% Additional control. E

% \begin{figure}[ht]
% % \vskip 0.2in
% % \begin{center}
% \centering
% \centerline{\includegraphics[width=\linewidth]{figures/fig2ter.png}}
% \caption{ We separated the previous statistics over the different recording slices, gathering voxels in each slice separately. Color indicate a different brain hemisphere and experiment (N or V indicate experiment name, A or T the animal name, R or L the right or left hemisphere). Most of the slices presented significant results across the 3 random repetitions of the in silicon training. {\color{red} add stats} }
% \label{fig:fig2ter}
% % \end{center}
% % \vskip -0.2in
% \end{figure}

In the above results, the Wav2Vec 2.0's responses to each sounds are strictly independent, because no back-propagation is being applied in between sounds. On the contrary, ferrets hear the sounds with a specific order, and may thus display brain responses that vary depending on this specific sound history. 

To test whether continuously-updating Wav2Vec 2.0 makes it more "brain-like", we compute brain scores on a "continuously back-propagated model" \emph{i.e.} a model that changes after the presentation of each sound.

The brain scores (F1: $r=0.0265 \pm 2\times10^{-4}$; F2: $r=0.0237 \pm 1.5\times10^{-4} $) of a continuously-updating model were significantly higher (F1: $p<10^{-7}$; F2: $p<10^{-8}$) than those of a frozen pretrained model (F1: $r=0.0212 \pm 1\times10^{-4}$; F2: $r=0.0173 \pm 1\times10^{-4}$, Figure \ref{fig:fig2}. A continuously updating model with a different order of sound presentation had similar brain score than the frozen pretrained model (F1: $r=0.0209 \pm 1\times10^{-4}$, $p=0.97$ ; F2: $r=0.01710 \pm 1\times10^{-4}$ , $p=0.99$). This effect was robust to the choice of order permutation, and present across all animals, hemisphere and most recording slices (Figure \ref{fig:fig3}).

\subsection{The improvement of brain-scores induced by back-propagation depends on the sound history}

\begin{figure}[ht]
% \vskip 0.2in
% \begin{center}
\centering
\centerline{\includegraphics[width=\linewidth]{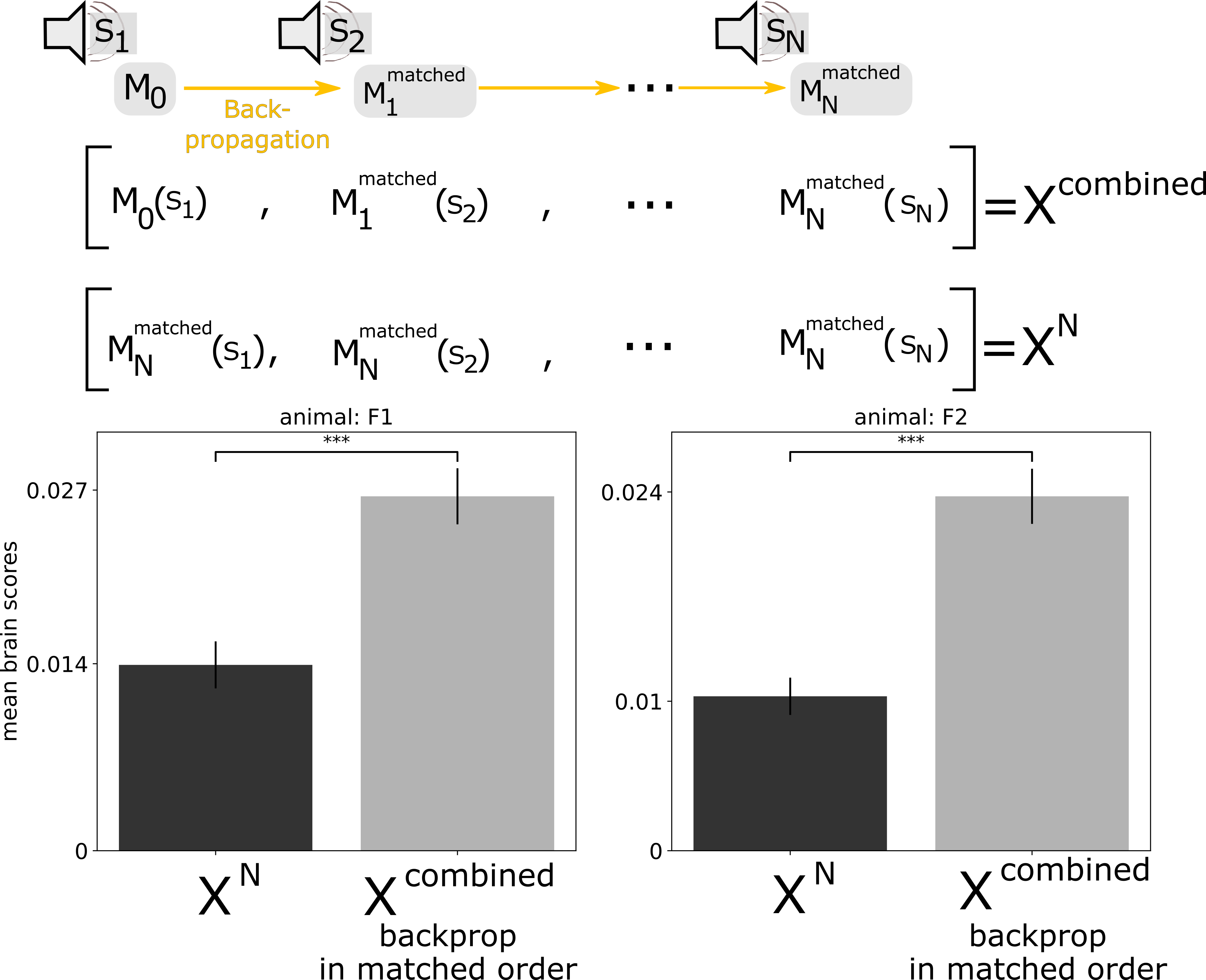}}
\caption{Network do not become more ferret-like. Brain scores of a post-training model $M_{N}$ are significantly smaller than brain scores of the continuously updating models. }
\label{fig:fig5}
% \end{center}
% \vskip -0.2in
\end{figure}

% Dynamics and controls. Dark line correspond to the early freezing experiment, while lighter line correspond to the control experiment. Sounds are ordered by their presentation order for each of the four repetition. We report results for the first slice that was recorded in each animal and hemisphere. In both case we repeat each in silico training 3 times and display the 3 trajectories.

Wav2vec 2.0 is originally pretrained with speech sounds only \citep{baevski_wav2vec_2020, panayotov_librispeech_2015}. In this experiment, however, we input this model with other types of sounds, including ferret vocalization and music. Wav2Vec 2.0 may thus fine-tune its parameter to model the specific sounds used in the present study, and one could hypothesize that this fine-tuning explains the increase of brain scores between a pretrained (no back-prop) model and a continuously-updated model.

Two controls infirm this hypothesis. First, 
%we showed above that the result was dependent on the exact order with which sounds are back-propagated
we evaluate whether updating Wav2Vec 2.0 with the order of a given session '$k$' to predict the brain responses recorded in different session '$k'$'. As displayed in Figure \ref{fig:fig4}, brain scores based on these mismatch orders are virtually identical to those obtained without back-propagation. 

Second, the brain scores obtained with a model fine-tuned to all the sounds (\emph{i.e.} $M_{N}$) , (F1: $r=0.0139\pm 10^{-4}$, F2: $r=0.0103\pm 10^{-4} $)  are significantly smaller than brain scores obtained with a continuously updated model (F1: $p<1.2\times10^{-7}$, F2: $p<1.2\times10^{-8}$), as well as brain scores obtained from a network without backpropagation, $M_{0}$, (F1: $p< 10^{-6}$, F2: $p<10^{-5}$, Figure \ref{fig:fig4}).

Therefore the increase of brain-score induced by back-propagation in Wav2Vec 2.0 is consistent with a quantitatively similar history-specific modulation of the brain responses.

\section{Discussion}

Recent studies have repeatedly shown similarities between brain activity and pretrained supervised neural networks (\emph{e.g.} \citep{yamins_performance-optimized_2014,yamins_using_2016,kell_task-optimized_2018,eickenberg_seeing_2017,millet_inductive_2021}). The present results further show that updating a model after each stimulus presentations \emph{via} self-supervised learning makes this model more "brain-like".

This phenomenon offers a computational framework to model plasticity in the brain. Indeed, acoustic context can impact perception and behavior \citep{agus_rapid_2010} through rapid plasticity neural mechanisms which store in memory the past sensory context. %Here we showed that trial-by-trial fluctuations of cortical responses are partly due to the preceding acoustic context. 
%We could demonstrate that what is traditionally considered as noise in hemodynamic responses constitutes in fact an early trace of cortical learning. 
% By definition, such mechanism cannot be captured by "frozen" artificial neural networks with no memory capacities, which is why we used a self-supervized network that can capture dynamics over long timescales. 
The most striking result of this study is that self-supervised back-propagation altered the deep neural network such that its activations better matched subsequent cortical responses.  

The convergence between rapid cortical plasticity and self-supervized backpropagation here faces several limits. First, voxels and artificial neurons are here averaged over time and summarize sound-evoked response to a single value because of the limited temporal resolution of fUS. Future work will thus be necessary to track the precise dynamics underlying our findings.
Second, the architecture and loss of Wav2Vec 2.0 is unlikely to identical to the brain's, whose loss is believed to be hierarchical and whose optimization is largely done locally. The use of a specific architecture, optimizer, self-supervised loss is only justified by the relative good performance this network achieved on speech recognition tasks. Further explorations and theoretical development are thus needed to identify which architecture, optimizer, training dataset and loss would provide the best predictor to the brain fluctuations here observed. 
Third, the brain scores achieved in the present study are small. This issue is pervasive across studies \citep{yamins_using_2016,millet_inductive_2021,caucheteux_gpt-2s_2021}, and results from the notoriously small signal-to-noise ratio of brain recordings. Interestingly, it is common to compute a "noise ceiling" analysis to evaluate the impact of noise. Noise-ceiling is a brain score computed, not from the activations of a model, but from the brain recordings of a repetition. By showing that the brain constantly changes as a function of previous stimuli, our study thus challenges noise-ceiling analyses based on repeated exposures.

% However one of the main reason for low brain score is due to the fact that fUS responses are intrinsically noisy. %If we had applied a full denoising procedure onto the fUS responses, the brain scores would have reached values centered around 0.6 since trial-by-trial fluctuations would be lost. 

Overall, the present study paves the way to understand and model the long-term modifications of biological and artificial neural networks, and their role in the optimization of an efficient self-supervised learning strategy.

\section*{Acknowledgments}
We would like to thank Agnès Landemard for her help concerning the dataset she recorded, as well as her many feedbacks.
YB was supported by the Fondation Fyssen, the FrontCog grant
ANR-17-EURE-0017, ANR-JCJC-DynaMiC, and SESAME fUS-PSL.
This  work  was  supported  by  ANR-17-EURE-0017,  the Fyssen Foundation and the Bettencourt Foundation to JRK for his work at PSL.
This project has received funding from the European Union’s Horizon 2020 research and innovation program under the Marie Skłodowska-Curie grant agreement No 945304.

\nolinenumbers

%This is where your bibliography is generated. Make sure that your .bib file is actually called library.bib
% \bibliography{library}
%This defines the bibliographies style. Search online for a list of available styles.

\setcitestyle{authoryear,round,citesep={;},aysep={,},yysep={;}}

\bibliography{main,self-supervisednet}
\bibliographystyle{icml2021}

\end{document}